%% file: main.tex
\documentclass[runningheads]{llncs}
\usepackage{graphicx}
\usepackage{multirow}
\usepackage{hyperref}
\usepackage{amsmath, bm}
\DeclareMathOperator*{\argmax}{arg\,max}

\usepackage{wrapfig}
\usepackage{booktabs} 

\usepackage{times}
\usepackage{xspace}
\newcommand{\sine}{SInE\xspace}

\usepackage[nocompress]{cite}

\usepackage[textsize=scriptsize
           ]{todonotes}

\begin{document}
\title{Graph Sequence Learning for Premise Selection}
%

%

\author{Edvard K. Holden
\and
Konstantin Korovin
}

\author{Edvard K. Holden
\and
Konstantin Korovin
}
\authorrunning{Holden \& Korovin}

\institute{The University of Manchester}

%

\maketitle              
\begin{abstract}

Premise selection is crucial for large theory reasoning as the sheer size of the problems quickly leads to resource starvation.
This paper proposes a premise selection approach inspired by the domain of image captioning, where language models automatically generate a suitable caption for a given image.
Likewise, we attempt to generate the sequence of axioms required to construct the proof of a given problem.
This is achieved by combining a pre-trained graph neural network with a language model.
We evaluated different configurations of our method and experience a 17.7\% improvement gain over the baseline.

\end{abstract}

\section{Introduction}

\input{introduction}

\section{Related Works}\label{sec:related_works}

\input{related_works}


\section{Axiom Captioning}\label{sec:captioning}

\input{method_caption}


\section{Problem Embeddings}\label{sec:problem_embedding}\label{sec:embedding}

\input{method_graph}


\section{Experimental Evaluation}\label{sec:evaluation}

\newcommand{\vocabsize}{vocab_6k}

\input{experiments}


\section{Conclusion}\label{sec:conclusion}

In this paper, we presented a novel approach for performing premise selection.
It parallels image captioning, combining transfer learning on graph neural networks with sequence learning.
The graph representation provides a holistic view of the problem structure, while the sequence model uses this embedding to predict the sequence of axioms necessary for the proof.
Our evaluation found that the model performs better when the GNN is pre-trained on a related and supervised task with embeddings containing information of all the nodes in the graph.
Further, we observed that effective axiom captioning requires a fixed axiom order and a greedy decoder sampler.
Lastly, the proposed approach dramatically increases the number of solved problems when complemented with \sine and significantly outperforms related machine learning methods.

\subsection*{Acknowledgements}

We thank Michael Rawson for providing a translator from formulas to graphs, utilised in this work.


%
%
%
%

\bibliographystyle{plain}
\bibliography{ref}

\end{document}

%% file: introduction.tex

Automated Theorem Provers (ATPs) construct formal proofs without human interaction and have seen great success in hardware and software verification, as it automatically verifies system properties. 
They have also played a significant role in formalisation projects such as Mizar~\cite{DBLP:journals/jar/Urban06} and as hammers in interactive theorem proving~\cite{DBLP:conf/lpar/PaulsonB10,DBLP:conf/cade/SteenB18}.

State-of-the-art ATPs such as iProver~\cite{DBLP:conf/cade/Korovin08,DBLP:conf/cade/DuarteK20}, E~\cite{DBLP:conf/lpar/Schulz13}, Vampire~\cite{DBLP:conf/cav/KovacsV13} and SPASS~\cite{DBLP:conf/cade/WeidenbachDFKSW09} attempts to solve problems consisting of a conjecture and a set of axioms through saturation.
This process consists of clausifying the input problem and computing all possible inferences between the clauses until it derives a contradiction or the set of clauses is saturated.
However, computing inferences quickly lead to a combinatorial explosion in the number of clauses which exhausts the computational resources.
Therefore, the proof search is guided by heuristics.
This is essential because the chance of deriving a contradiction reduces considerably as the number of clauses grows.
Machine learning is currently being used to discover efficient heuristics~\cite{DBLP:conf/gcai/SchaferS15, DBLP:journals/corr/JakubuvU16a, DBLP:conf/mkm/HoldenK21}, and to intelligently operate internal heuristic components~\cite{DBLP:conf/cade/Bartek021, DBLP:conf/frocos/GoertzelCJOU21, DBLP:journals/jar/FarberKU21, DBLP:conf/cade/RawsonR18}.

However, strong heuristics are insufficient when reasoning over large theories.
Problems concerning verification and mathematical statements often contain a vast amount of axioms,  quickly resulting in resource starvation due to the sheer size of the initial space.
A key observation is that, despite the large set of axioms, only a fraction is typically required to construct the proof.
Consequently, by removing all ``superfluous'' axioms, the problems become computationally feasible, and the chance of finding proof increases dramatically.
This task is known as \textit{premise selection}.

This paper explores the adaptation of image captioning to the premise selection task.
Image captioning models aim to produce a sentence in natural language that describes a given image.
The captions are generated by embedding images using a pre-trained image model and combining the embedding with a language model.
Such methods are novel for premise selection and hold multiple compelling properties.
First, 
the axioms are represented by tokens, and their embeddings are learnt during training.
With an abstract token representation of the axioms, we can leverage both the conjecture-axiom and the inter-axiom relationship.
This 
 is in contrast to
approaches that accentuate the structural and symbol similarity of conjecture-axiom pairs.
A vital property of the language model is encapsulating sequences of axioms as opposed to treating them separately.
By representing the axioms occurring in a proof as a sorted set, the model can learn the conditional relationship between the axioms occurring in a proof.
This is crucial for treating the axioms as a collective set.

Another challenging aspect of a captioning approach is computing problem embedding entailing problem semantics.
First-order problems consist of a set of tree-structured formulas which are not easily represented through a feature vector, as required for machine learning.
This paper investigates pre-trained graph neural networks (GNNs) to embed problems via transfer learning.
GNNs operate on graphs and can incorporate structural properties into the embedding.
Nevertheless, there is no apparent pre-training task for FOF problems as there is for, e.g. images.
Therefore, we investigate using a supervised pre-training step where the GNN learns the embedding by training on the premise selection problem in the binary setting.
Additionally, we experiment using an unsupervised approach that learns to embed the problems based on graph similarity.
The graph embeddings are further enhanced by emphasising different sections of the embedding at given steps of axiom generation by exploring attention mechanisms.

Due to the challenges of premise selection, a single approach is unlikely to encapsulate all productive aspects of the axiom-conjecture relationship.
Hence, we also explore the combination of our method and SInE~\cite{DBLP:conf/cade/HoderV11}.

Our main contributions are:
\begin{itemize}

    \item Adapt approaches from image captioning to the task of premise selection.

    \item Novel method for combining graphs with sequence learning in the context of premise selection, outperforming previous tokenised conjecture approaches.
    
    \item Novel method for unsupervised training of GNNs embeddings of FOF problems.
    
    \item Usage of GNNs on problem graphs for transfer learning.
    
    \item 'Rare Axiom' inclusion technique for training with a reduced vocabulary while maintaining rare positive axioms.


\end{itemize}

We evaluated our approach over an extended version of the DeepMath dataset.
The results show that the specificity of our approach, in combination with the breadth of \sine, significantly outperforms related methods and results in a 17.7\% increase in the number of solved problems over the baseline.

This paper is structured as follows: in Section~\ref{sec:related_works} we present the related works.
In Section~\ref{sec:captioning} we present the sequence model and in Section~\ref{sec:problem_embedding} we describe our approach for obtaining problem embeddings.
In Section~\ref{sec:evaluation}, we evaluate our approach experimentally both offline and online, before concluding the paper in Section~\ref{sec:conclusion}.

%% file: related_works.tex
Premise selection has previously been addressed with heuristic-based methods such as MePo~\cite{DBLP:journals/japll/MengP09} and the very successful \sine~\cite{DBLP:conf/cade/HoderV11} algorithm.
The core idea of \sine is that axioms are likely to contribute towards the proof if they contain symbols related to the symbols in the conjecture.
This is achieved by iteratively selecting axioms with symbols occurring in a set of selected axioms relative to how often the symbol appears globally. 
The main limitation of the approach is a low specificity and not utilising any information from existing proofs.

The task of premise selection has also been approached with machine learning methods such as Naive Bayes~\cite{DBLP:journals/jar/Urban06}, kernel methods~\cite{DBLP:journals/jar/AlamaHKTU14}, K-NN~\cite{DBLP:journals/jar/KaliszykU15a} and binary classification~\cite{DBLP:journals/corr/AlemiCISU16, DBLP:journals/corr/abs-1807-10268,  https://doi.org/10.48550/arxiv.1802.03375, DBLP:conf/cade/RawsonR20}.
In the binary setting, the goal is to train a supervised model to score conjecture-axiom pairs.
A significant drawback of this method is that axioms are considered independent, and the problem sizes strongly skew predictions.
Instead, axioms should be treated as a collective entity, as all the axioms occurring in a proof must be selected to construct the proof.

The approaches most similar to our method are the sequence-to-sequence approach in~\cite{DBLP:conf/lpar/PiotrowskiU20} and its extension with a Transformer model~\cite{DBLP:conf/mkm/ProrokovicWS21}.
The sequence models treat the conjecture as a sequence of tokens and map it to a sequence of axioms.
Their main limitation is being unaware of how the conjecture relates to elements of the various axioms.
GNNs can model the relationship between formulae elements, as shown by the binary graph classification approach in~\cite{DBLP:conf/cade/RawsonR20}.
Meanwhile, our approach is aware of the relationship between the axioms occurring in the proof and the conjecture's connection to these axioms.


%% file: method_caption.tex
The image captioning problem can be stated as follows: given an image $I$ and a dictionary of words $\Omega$, generate an accurate and grammatical caption $S$, consisting of words from $\Omega$.
This challenging problem goes beyond the already non-trivial task of identifying the image objects.
Rather, it requires identifying and comprehending: the objects, their attributes and their relation.
Moreover, this information must be decoded and represented as a grammatically correct sentence in the target language.

\begin{figure}
\centering
\includegraphics[scale=0.38]{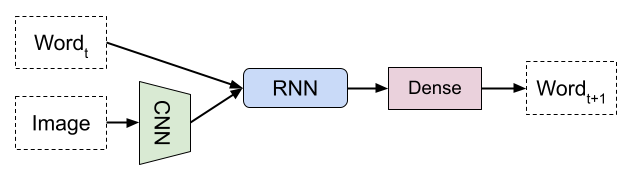} 
\caption{The inject architecture for image captioning.}
\label{fig:captioning_model}
\end{figure}

State-of-the-art image captioning approaches join the machine learning fields of image classification and language modelling.
An example of a captioning model based on the inject architecture is shown in Figure  \ref{fig:captioning_model}.
It consists of three components: an image encoder, a language model, and a dense output layer.
The image encoder extracts and embeds the image semantics as a feature vector.
The language model combines these salient features with an input word to produce an encoding of the current sequence.
Finally, the dense layer maps the encoding to a probability distribution over the vocabulary.

Despite the challenges of image captioning, the models produce appropriate and detailed image descriptions.
Due to their expressiveness, we believe these methods can be utilised for premise selection.
In the remaining parts of this section, we describe the sequence model.


\subsection{Sequence Learning}

In the original task of image captioning, the model operates on pairs of images and captions in a target language.
In the context of premise selection, the images are replaced by problems and the captions are replaced with the axioms that appear in the proof of the problems.
Assume we have a problem $I$ with a corresponding proof $S^*$ and an axiom resource bank $\Omega$.
Next, we extract and impose an order on the $m$ axioms in $S^*$, resulting in $S =\langle s_1, \ldots , s_m \rangle$, $s_i \in \Omega$ for $1\leq i \leq m$. 
We describe the task of premise selection in the context of sequence learning as maximising the probability of producing the sequence of axioms used in the proof of a given problem.
Given the problem-axioms pair $(I, S)$ we can compute its log probability as:

$$
\log p (S | I) = \sum_{t=1}^m \log p(s_t | s_{t-1}, \ldots, s_1, I).
$$

We estimate $\log p(s_t|s_{t-1},\cdots, s_1, I)$ with the recurrent neural network (RNN) $\sigma$ with learnable parameters $\theta$.
RNNs exhibit a dynamic behaviour over a sequence of inputs due to their internal memory state $h_t$, which captures the previous inputs sequentially.
In particular, the output at step $t$ depends on the previous memory state $h_{t-1}$ and the input $s_{t-1}$.
The hidden state is defined as:

\[
h_t = 
\left\{
\begin{array}{r@{\quad}l}
     \sigma(I;\theta) & \text{if } t = 1,\\
    \sigma(h_{t-1}, s_{t-1};\theta)  & \text{otherwise}.\\
\end{array} \right.
\]

The RNN is trained to predict the next token in a sequence based on the previous token and the current memory state.
Over a training set of problem-axiom pairs  $\{(I^i, S^i)\}^N_{i=1}$, the model is trained to maximise the log probability of producing the correct sequence of axioms:
$$
\theta^* = \argmax_\theta \sum_{I,S} \log p(S|I;\theta).
$$

Thus, the model predicts axioms based on the problem and the previously predicted axioms.
In our implementation, we use Long-Short Term Memory (LSTM)~\cite{lstm} cells as the underlying RNN.
LSTM is among the most popular RNN models due to its robustness towards vanishing and exploding gradients.


\subsection{Axiom Captioning}

\begin{figure}[!htpb]
\centering
\includegraphics[width=0.95\linewidth]{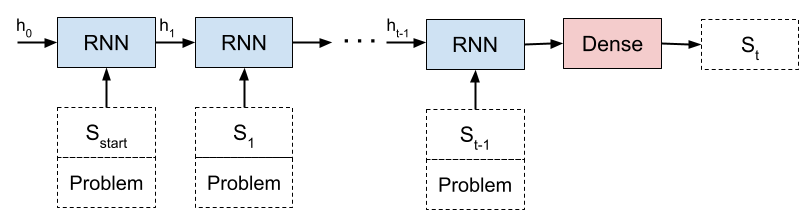} 
\caption{Recurrent Neural Network predicting the next token in a sequence.}
\label{fig:rnn_sequence}
\end{figure}

The generative axiom prediction model is constructed using the par-inject architecture \cite{DBLP:journals/corr/TantiGC17}, as illustrated in Figure \ref{fig:rnn_sequence}.
This architecture takes a token embedding $\textbf{s}$ and a problem embedding $\textbf{I}$ at each time step.
The model is given the special start token $s_{start}$ to initialise the axiom generation process.
Likewise, a special end token, $s_{end}$, represents the end of a sequence. 
Consequently, start and end tokens are added to each axiom sequence such that the model is trained on the target sequence  $\langle s_{start} , s_1, \ldots , s_m, s_{end} \rangle$.
Axioms with few occurrences in the dataset are replaced by the Out-Of-Vocabulary token $s_{unkown}$.
These three special tokens are included in the dictionary $\Omega$.


At training time, we apply teacher forcing., which feeds the next token of the training sequence to the model instead of its previous prediction.
This prevents the model from being unable to recover from poor predictions.
Hence, the prediction at each training step is expressed as:

\[
\hat{y_t} = 
\left\{
\begin{array}{r@{\quad}l}
      \sigma( \textbf{s}_{start}, \textbf{h}_{0}, \textbf{I} ; \theta ) & \text{if } t = 1,\\
    \sigma(\textbf{s}_{t-1}, \textbf{h}_{t-1}, \textbf{I} ; \theta)  & \text{otherwise.}\\
\end{array} \right.
\]

Where $\hat{y_t}$ is a probability distribution at time $t$ over the axioms in $\Omega$.


\subsection{Neural Attention Captioning} \label{sec:attention}

The captioning decoder is fed a static input entity at each time step, but it can be advantageous to emphasise different parts of the embedding based on the current model state~\cite{https://doi.org/10.48550/arxiv.1502.03044}.
This is achieved through a separate attention network that dynamically weighs some input according to a model state.
In other settings, the attention mechanism can emphasise particular words in a sequence or regions of an image with respect to the model state.
In this scenario, the incentive of attention is to emphasise particular sections and elements of the averaged graph representation to enhance the embedding.

The attention mechanism computes a context vector which is used as input to the next stage of the model.
It is a weighted sum of the $n$ embedding elements where each weight is the quantity of attention applied to the corresponding element:

\[
\bm{c}_t = \sum_{i=1}^n \alpha_{t,i} \bm{I}_i.
\]

The weights $\alpha_{t,i}$ are computed based on an alignment score function which measures how well each element matches the current state.
The scores are scaled by softmax into weights in the range of $[0, 1]$ where the sum of the weights equals to 1:
\begin{wrapfigure}[14]{}{0.35\textwidth}
\includegraphics[width=0.98\linewidth]{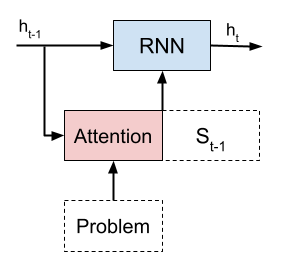} 
\caption{Language model with Bahdanau attention.}
\label{fig:attention}
\end{wrapfigure}

\[
\alpha_{t,i} = 
\frac{
\exp(score(\bm{I}_i, \bm{h}_t))
}
{
\sum_{j=1}^n \exp(score(\bm{I}_j, \bm{h}_t))
}.
\]

In Section~\ref{exp:attention}, we experimented with both Loung attention~\cite{DBLP:journals/corr/LuongPM15} and Bahdanau attention~\cite{https://doi.org/10.48550/arxiv.1409.0473}.
The alignment function of all three attention types is shown in Table~\ref{tab:attention_alignment_functions}, where 
$W, W_1, W_2$ and $V$ are learnable attention parameters.
In the Bahdanau style, the context vector is concatenated with the token embedding and fed to the RNN decoder, as illustrated by Figure~\ref{fig:attention}.
This is a key difference from the Loung style, where the alignment scores are computed on the output of the RNN prior to the dense layer.

\begin{table}[!htpb]
    \centering
    \begin{tabular}{@{}l c l l @{}} 
         Attention Style & &  & Alignment  \\ \toprule
         Bahdanau       & &  score($\bm{I_i}, \bm{h}_{t-1}$)  & = $V^\top \cdot tanh( W_1 \cdot \textbf{I}_i + W_2 \cdot \textbf{h}_{t-1}) $    \\ 
         Loung Dot      & &  score($\bm{I_i}, \bm{h}_{t}$)    & = $\bm{h}_{t}^\top \cdot \bm{I}_i  $ \\
         Loung Concat   & &  score($\bm{I_i}, \bm{h}_{t}$)    & = $V^\top \cdot tanh( W[\textbf{I}_i;\textbf{h}_{t}]) $   \\ \bottomrule
    \end{tabular}
    \caption{Overview of attention alignment functions.}
    \label{tab:attention_alignment_functions}
\end{table}


%% file: method_graph.tex
An embedding is a fixed-size, real-valued vector representation of an entity, where semantically similar entities ideally are close in the embedding space.
In the original task of image captioning, image embeddings consist of low-level image features obtained from pre-trained convolutional neural networks over extensive image classification datasets.

Computing problem embeddings in a similar fashion pose multiple challenges in the context of first-order problems.
Firstly, the problems have no natural fixed-size vector representation as they consist of unordered sets of tree-structured formulae.
Thus, encoding syntactic, structural and semantic properties as a vector is non-trivial.
Secondly, there is no immediate classification task for learning the semantics of first-order problems.
This paper attempts to overcome these challenges by producing embeddings via graph neural networks.

\subsection{Problem Graph}

\begin{figure}[!htpb]
    \centering
    \begin{minipage}{0.45\linewidth}
        \centering
        \includegraphics[width=0.95\linewidth]{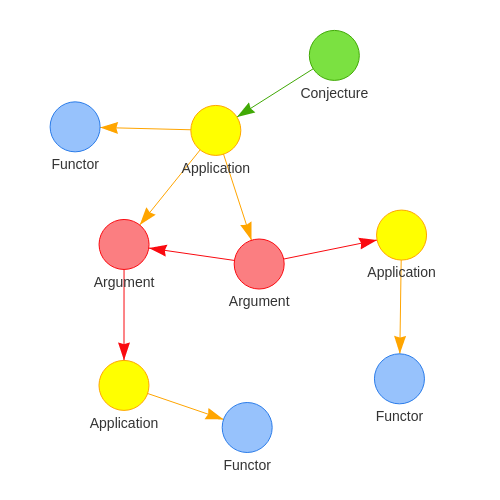}
        \caption{A conjecture graph.}

        \label{fig:graph_conjecture}
    \end{minipage}\hfill
    \begin{minipage}{0.45\linewidth}
        \centering
        \includegraphics[width=0.95\linewidth]{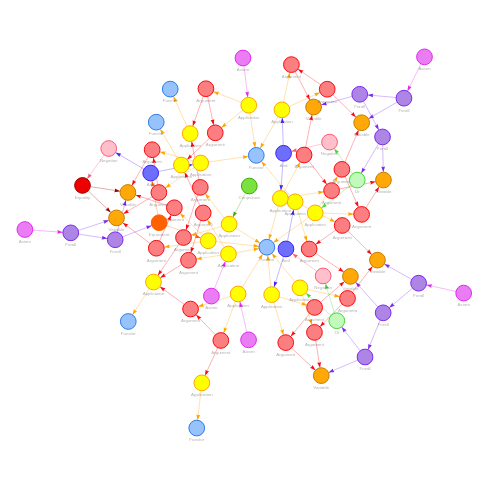} 
        \caption{A problem graph of one conjecture and six axioms.}
        \label{fig:graph_problem}
    \end{minipage}
\end{figure}

A first-order logic formula has an intrinsic tree-shaped structure and is naturally represented as a directed acyclic graph $G$ with vertices $V$ and edges $E$.
The vertices, also known as nodes, correspond to the types of elements occurring in the formula, such as function symbols and constants.
The edges denote a relationship between the vertices, e.g., an argument supplied to a function.
Figure~\ref{fig:graph_conjecture} illustrates the graph representation of a conjecture, spanning four different node types as visually represented by the colouring. 

This representation extends to sets of formulas by computing a global graph over the node elements in the formulae, as shown in Figure \ref{fig:graph_problem}.
The graph representation captures many aspects of the formulae while invariant to symbol renaming and encoding problems with previously unseen symbols.
This paper uses a graph encoding of 17 node types as described in~\cite{DBLP:conf/cade/RawsonR20}.


\subsection{Graph Neural Networks}

The problem graph is embedded into an $n$-dimensional embedding space via a graph neural network.
A graph neural network is an optimisable transformation that operates on the attributes of a graph.
It utilises a ``graph-in, graph-out" methodology where it embeds the graph while preserving the structure and connectivity of the original graph.

A randomly initialised vector represents each node type $\Phi$ across all graphs in an $n$-dimensional embedding space.
Next, each node in a graph is assigned to its corresponding embedding vector $\bm{x}_\Phi$, resulting in the node feature matrix $X$.
The GNN embeds the type features of each node $\bm{x}_\Phi$ into the node feature embedding $\bm{x}_\Phi'$ through a node update function.
This effectively transforms the graph features $X$ into a more favourable embedding  $X'$.
Adjacent nodes are incorporated into the update of a node to encode the structure through message passing~\cite{DBLP:journals/corr/GilmerSRVD17}.

Message passing is accomplished through graph convolutional layers, and we utilise the operation described in~\cite{https://doi.org/10.48550/arxiv.1609.02907}.
The node-wise convolutional operation for the attributes $\bm{x}_i^{(k)}$ of node $i$ at step $k$ is described as:

\[ 
\bm{x}_i^{(k)} = W
\sum_{j \in \mathcal{N}(i) \bigcup \{i\}}
\frac{e_{j,i}}
{\sqrt{ \hat{d_j} \hat{d_i} } }
\bm{x}_j^{k-1}
\]	

where $W$ is a learnable weight matrix, $\mathcal{N}(i)$ is the set of neighbouring nodes of $i$ and $\hat{d_i} = 1 + \sum_{j \in \mathcal{N}(i)}e_{j,i}$. 
The variable $e_{j,i}$ denotes the edge weight from $j$ to $i$.
 In this setting, all edge weights are 1. 
The convolutional operations are applied synchronously to all nodes in the graph and learn hidden layer representations that encode both local graph structure and nodes features.




\begin{figure}
\centering
\includegraphics[scale=0.38]{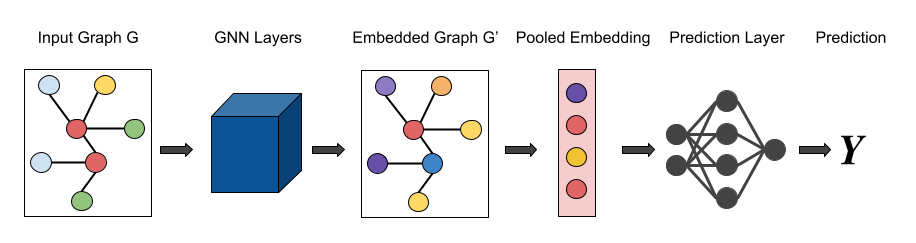} 
\caption{Graph Neural Network for classification of graph or node properties.}
\label{fig:graph_workflow}
\end{figure}

After computing the graph embeddings, they are pooled and passed through the prediction layer, which produces the final model output.
We experiment with three different mean pooling approaches all nodes in the graph, only axiom nodes, and only the conjecture node.
An overview of the GNN pipeline is shown in Figure~\ref{fig:graph_workflow}.

In this approach, the GNN is pre-trained on auxiliary tasks, computing the embeddings before training the captioning model.
We experiment with supervised and unsupervised pre-training GNN approaches, as described below.

\subsection{Supervised Problem Embedding}

In the supervised approach, the GNN is trained on the node level by performing binary premise selection over the axiom nodes, as described in~\cite{DBLP:conf/cade/RawsonR20}.
Based on their node embedding, the model learns to predict whether an axiom occurs in the proof of a problem.
During training, the resulting axiom node embeddings become increasingly valuable for modelling their contribution towards the proof.
Therefore, the node embeddings are expected to contain information crucial to premise selection.
Our experiments show that this information prevails through average pooling.



\subsection{Unsupervised Problem Embedding}

The supervised learning task emphasises the axiom nodes, but it might be advantageous with a learning task encapsulating all the nodes in a graph.
Alas, no sensible labels are directly derivable from the problems to train a prediction model.
Thus, we employ unsupervised training through a synthetic dataset which utilises a relation property encapsulating all graph nodes.

\begin{figure}[!htpb]
\centering
\includegraphics[scale=0.37]{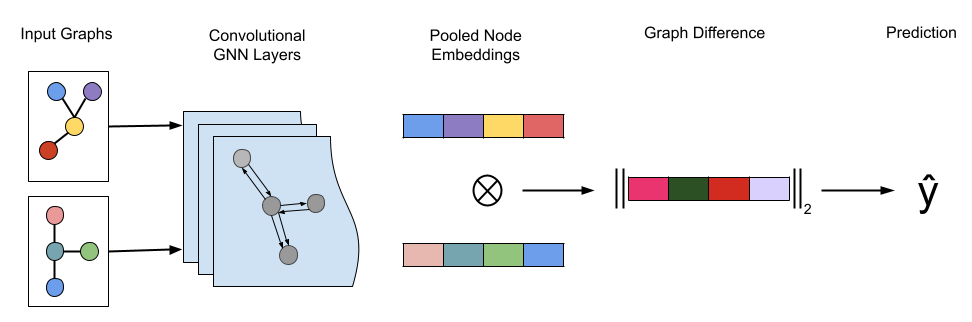} 
\caption{Unsupervised GNN training based on pairwise graph similarity.}
\label{fig:embedding_unsupervised}
\end{figure}

The unsupervised training approach consists of training a matching model which learns the difference between two graphs according to some relational property, as described in~\cite{unsupervised_graph_classification}.
The model takes two graphs, $g_i$, $g_j$, as input and passes them through the Siamese GNN model, as illustrated in Figure \ref{fig:embedding_unsupervised}.
Next, the nodes of the embedded graphs are pooled into two graph embedding vectors.
The similarity of the two input graphs is approximated as the vector norm between the two graph embeddings:  $ || GNN(g_i) - GNN(g_j) || $. 
Training the GNN in this fashion enables it to produce embeddings encompassing structural similarities and dissimilarities.

The synthetic dataset consists of pairs of undirected graphs and a numeric property describing their relation.
The relational property utilised is the Laplacian spectrum distance~\cite{Wills_2020}, which can be defined as follows.
Given a  graph $G$, the adjacency matrix $A$ represents the node connections in the graph.
The diagonal degree matrix $D$ of $G$ represents the degree of each node, e.g. the number of neighbours.
Further, the Laplacian of the graph is defined as the degree matrix subtracted from the adjacency matrix:

\[
L = D - A
\]

\noindent
The eigenvalues $\lambda_1 \leq \ldots \lambda_i \ldots \leq \lambda_k$ of the Laplacian are given as $L \bm{x} = \lambda_i \bm{x}$.
Accordingly, the Laplacian spectrum distance $\pi$ of two graphs $G$ and $G'$, is defined as:


\[
\pi (G, G') = 
\sqrt{
\sum_{i=1}^{k} (\lambda_i - \lambda_i')^2
},
\]
where $k \in min(n, m)$, and $n$ and $m$ are the numbers of nodes in $G$ and $G'$.
 
The Laplacian spectrum distance is a computationally cheap metric, even for graphs of the magnitude required to represent first-order problems.
Although the metric encapsulates graph structure, it neither considers node types nor edge directions. 
Still, the distance provides an overall description of the structural similarity of the graphs and considers all graph nodes. 

%% file: experiments.tex
This section describes the experimental results and evaluation of our premise selection approach\footnote{Experiments available at \url{https://github.com/EdvardHolden/axiom\_caption}}.
First, we describe the dataset, followed by five experiments.
The first experiment investigates the performance of the graph embeddings\footnote{Embedding computation available at \url{https://github.com/EdvardHolden/gnn-entailment-caption}}.
The second experiment explores input orders, and the third investigates the effect of attention.
The fourth experiment explores decoder sampling methods, and the fifth is an online evaluation of our approach and related methods.


\subsection{Proof Dataset}

We used the synthetic DeepMath~\cite{DBLP:journals/corr/AlemiCISU16} dataset for the model training and evaluation.
It consists of 32524 FOF problems based on proofs of the Mizar40 benchmark.
DeepMath was created to evaluate binary premise selection methods.
Hence, the number of positive and ``superfluous'' axioms in a problem is balanced.
The main advantages of the dataset are a large number of problems combined with consistent formula naming and the reuse of axioms across problems.

We impose a maximum sequence limit of 20 axioms resulting in 30805 problems, where 20\% are used for testing, 8\% for validation, and the rest for training.
The vocabulary consists of the 6K most common axioms occurring in the proofs of the training set.
The other axioms are mapped to the Out-Of-Vocabulary (OOV) and removed.
Table~\ref{table:dataset} displays the key statistics of each problem set.
While many proofs contain rarely occurring axioms, few problems are represented solely by OOV tokens.



\begin{table}
\centering

\input{results/\vocabsize/problem_statistics}

\caption{Key statistics of each dataset partition.}
\label{table:dataset}
\end{table}


\subsection{Experiment 1: Supervised vs Unsupervised Graph Embeddings}

The goal of this experiment is to examine the pre-trained embedding methods.
The supervised approach was trained according to the methods in the original paper~\cite{DBLP:conf/cade/RawsonR20}.
The unsupervised approach was trained on 12000 graph pairs.
The output of each GNN is combined with one of the pooling mechanisms to produce a total of six embedding variations.
The captioning training parameters are shown in Table~\ref{tab:training_setup}.
The evaluation metrics are defined as follows, where $A$ is the set of predicted tokens, and $B$ is the ground truth:

\begin{align*}
Jaccard(A,B) = \frac{|A \cap B |}{| A \cup B |}  && Coverage(A, B) = \frac{|A \cap B |}{|B|} 
\end{align*}

\begin{table}
\centering
\begin{tabular}{@{}lllll@{}}
\multicolumn{2}{c}{\bf Training Prameters} &  & \multicolumn{2}{c}{\bf Model Parameters}     \\ \toprule
Optimiser               & Adam  &  & RNN Type              & LSTM  \\
Learning Rate           & 0.001 &  & RNN Units             & 32    \\
Max Epochs              & 80    &  & Stateful RNN          & True  \\
Early Stopping          & 5     &  & Batch Normalisation   & False \\ \midrule
Dropout Rate            & 0.1   &  & Feature Normalisation & True  \\
Teacher Forcing Rate    & 1.0   &  & No Dense Units        & 512   \\
Batch Size              & 64    &  & Embedding Size        & 50    \\
Maximum Sequence Length & 22    &  & Target Vocab Size     & 6000  \\ \bottomrule
\end{tabular}
\caption{The captioning model and training parameters.}\label{tab:training_setup}
\end{table}


The results are presented in Table \ref{table:graph_embedding_results} and show that supervised embeddings perform better than unsupervised embeddings.
This indicates that a contextual learning task is required to produce good embeddings.
It also shows that the embeddings created by pooling problem and premise nodes perform better, indicating that essential structures persist through averaging.
The results show issues with overfitting, but an increase in the dropout rate led to a decrease in validation performance.
On the other hand, unsupervised learning is relatively less prone to overfitting.
The following experiments utilise the supervised problem embeddings.



\begin{table}[!htpb]
\centering

\input{results/\vocabsize/experiment_features}

\caption{The training and validation performance of the embedding approaches.}
\label{table:graph_embedding_results}
\end{table}

\subsection{Experiment 2: Axiom Order}\label{exp:axiom_order}

This experiment examines various input ordering schemes.
While the ATP treats the axioms as a set, RNNs operate over input sequences.
Therefore, some input orders may be more advantageous.
We explore the following ordering schemes:

\begin{itemize}
    \item \textbf{Original}: Ordered as in the original problem.
    \item \textbf{Length}: Ordered the smallest string representation to the longest.
    \item \textbf{Frequency}: Ordered from the most frequently occurring axioms to the least. 
    \item \textbf{Random}: Random order of axioms for each sequence.
    \item \textbf{Global Random}: Static random order.

\end{itemize}

The results are displayed in Table~\ref{table:order_results}.
Although most configurations perform similarly, the length and frequency order have the best validation performance.
However, random ordering has a surprisingly low performance.
As this is not reflected in the other ordering schemes, including the global random order, it indicates that a consistent relative position across sequences is essential.

\begin{table}
\centering

\input{results/\vocabsize/experiment_order}

\caption{The train and validation set performance on the different axiom orders.}
\label{table:order_results}
\end{table}

\subsection{Experiment 3: Attention Mechanisms}\label{exp:attention}

This experiment evaluates the impact of utilising an attention mechanism for the axiom captioning method,
The models were trained with the length order, and the results are presented in Table~\ref{table:attention_results}.
The results show that Loung dot attention can provide a minor performance improvement.

\begin{table}
\centering
\begin{tabular}{lllllll}
\multirow{2}{*}{Attention} &  & \multicolumn{2}{c}{Train}     &  & \multicolumn{2}{c}{Validation} \\ \cline{3-4} \cline{6-7} 
                           &  & Jaccard       & Coverage      &  & Jaccard        & Coverage      \\ \cline{1-1} \cline{3-4} \cline{6-7} 
None                       &  & \textbf{0.43} & \textbf{0.53} &  & \textbf{0.24}  & 0.33 \\ 
Bahdanau                   &  & 0.37          & 0.50          &  & 0.22           & 0.32          \\
Loung concat               &  & 0.38          & 0.49          &  & \textbf{0.24}  & 0.33          \\
Loung dot                  &  & 0.39          & 0.51          &  & \textbf{0.24}  & \textbf{0.34} \\ \hline
\end{tabular}

\caption{The performance of attention mechanisms and no attention.}
\label{table:attention_results}
\end{table}


\subsection{Experiment 4: Model Decoding}

This experiment shifts the focus from optimising the training parameters toward the premise selection task by examining sampling methods.
Although the model is given a single input token at each step, multiple tokens can be sampled from the output distribution.
We explore three different sampling methods over the test set for selecting which axioms to include in the final problem:
\begin{itemize}
    \item \textbf{Greedy}: Select $n$ axioms with the maximum probability.
    \item \textbf{Top-K}: Redistribute the probability distribution over the top $K$ axioms, and sample $n$ axioms from the new probability distribution.
    \item \textbf{Temperature}: Scale the logits by the temperature prior to applying softmax and sample $n$ axioms from the new probability distribution.
\end{itemize}

At each generative step, the sampling method selects $n$ axioms and adds them to the set of the selected axiom.
Only the top axiom is fed back into the model.
The results are displayed in Table \ref{table:decoding_results}.
They show that the greedy sampling method is superior, and the coverage score is monotonic on $n$. We found that an increase in coverage gives a substantial improvement for online performance compared to a slight decrease in Jaccard. Moreover, greedy sampling selects a small total number of axioms despite producing the highest coverage scores.
Through experimentation, we discovered that achieving a high coverage score is crucial for solving the problems in this dataset.

\begin{table}
\centering

\input{results/\vocabsize/experiment_decoding}

\caption{The performance of each sampling method.}
\label{table:decoding_results}
\end{table}

\subsection{Experiment 5: Online System Evaluation}

The final experiment performs an online evaluation of our premise selection method and other methods with the state-of-the-art ATP, iProver\footnote{iProver is available at:\url{https://gitlab.com/korovin/iprover}}.
During initial experimentation, it was discovered that the Deepmath problems were too small for a meaningful assessment as iProver solved 86\% of the problems without premise selection. 
Therefore, the problems were extended by merging the Deepmath problems with the larger Mizar40\footnote{Problems are available at \url{http://grid01.ciirc.cvut.cz/~mptp/7.13.01_4.181.1147/mptp/problems_small_consist.tar.gz}} problems, reducing this ratio to 52\%. 


Nine different premise selection configurations are utilised to generate problem versions used in the online evaluation.
This consists of axiom captioning, SInE, no premise selection, and three related axiom captioning methods.
SInE is evaluated with tolerance-depth parameters of $(1, 1)$ and $(3, 0)$, note that $0$ corresponds to unbounded depth.

In the case of axiom captioning, the model is trained on the 6K most common positive axioms in the benchmark.
The positive axioms outside the vocabulary are essential for the proofs but appear too rarely to be learnt by the model.
Thus, these axioms are not reachable for the captioning model.
Consequently, our axiom captioning method implements a ``rare-axiom'' procedure for selecting rare but valuable axioms.
The procedure scans a given problem and selects any axioms that have previously appeared positively but rarely.
Furthermore, it is unlikely that a single method can encapsulate all aspects of the conjecture-axiom relationship.
Hence, we also evaluate the combination of axiom captioning and SInE as two complementing approaches.

\begin{figure}

 

\centering
\includegraphics[width=0.8\textwidth]{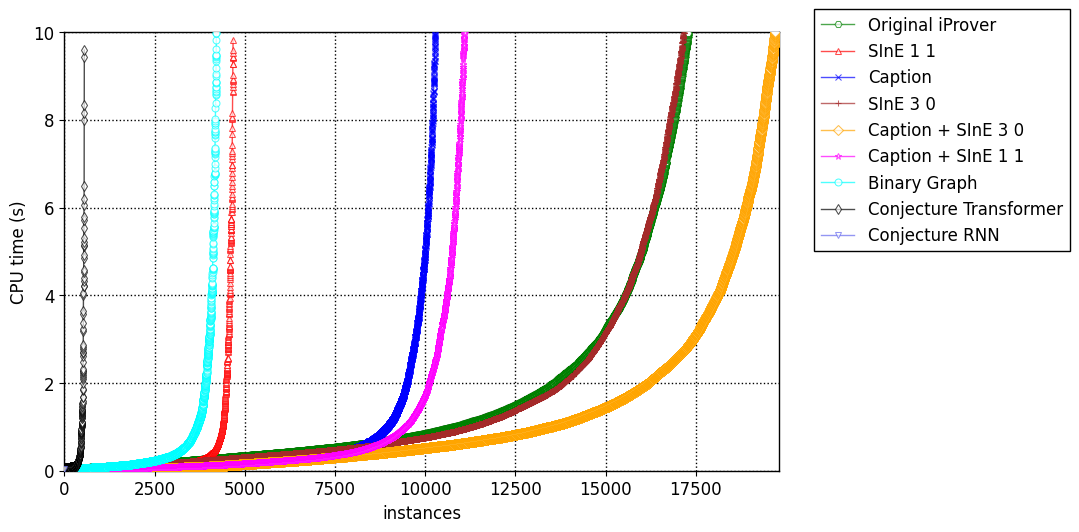} 
\caption{Online evaluation of SInE, Captioning, their combination, and related methods.}
\label{fig:cactus}
\end{figure}

Further, we evaluate three related ML-based methods for premise selection: Binary Graph ~\cite{DBLP:conf/cade/RawsonR20}, Conjecture RNN~\cite{DBLP:conf/lpar/PiotrowskiU20} and the Conjecture Transformer~\cite{DBLP:conf/mkm/ProrokovicWS21}.
The Binary Graph method is used to compute the supervised embeddings, and the comparison will therefore reveal whether axiom captioning improves the utilisation of the graph embeddings.
Still, the method expects a balanced dataset and will likely introduce much positive bias for larger problems.
Conjecture RNN is a sequence-to-sequence approach where the input is a tokenised embedding of the conjecture.
A caveat with this approach is that it trains on a specific axiom order derived from the proofs unavailable for this dataset.
Hence, it is trained on the consistent ``length" order (see, Section~\ref{exp:axiom_order}).
The approach utilises an LSTM architecture but puts no restriction on the input and output vocabulary, resulting in poor training performance on this dataset.
The Conjecture Transformer approach is an extension where the RNNs are replaced with the Transformer~\cite{https://doi.org/10.48550/arxiv.1706.03762} architecture.
This approach was modified slightly in the input and output layers to enforce input length restrictions and OOV tokens.
The Conjecture* approaches rely solely on the token vocabulary and do not inspect the given problem.
Hence, their encoders are unaware of the increased problem size.

iProver is run with a default heuristic over the clausified problem versions generated by each configuration with a time limit of 10 seconds.
The results are displayed in Figure~\ref{fig:cactus}.
Conjecture RNN only solves one problem selecting only the most frequently occurring axiom.
The ``Conjecture Transformer'' slightly improves the performance but does not come close to axiom captioning.
This suggests that graph embeddings are superior to conjecture token embeddings for premise selection.
Binary Graph performs best of the related methods and is close to the performance of SInE $(1,1)$.
However, it suffers from generating large problems with many superfluous axioms and a low coverage score due to being a binary classifier trained on a balanced dataset.
Nevertheless, half of the problems solved by Binary Graph complement the solution of axiom captioning.
Hence, graph embeddings facilitate method diversification.

Axiom caption outperforms the related ML-based methods and is slightly enhanced by its combination with SInE $(1, 1)$.
However, it is behind SInE $(3, 0)$, which performs nearly identically to no premise selection (Original).
The real power of axiom captioning comes through when paired with SInE $(3, 0)$.
Axiom captioning considers axioms carefully and consequently predicts many of the essential axioms.
On the other hand, SInE $(3,0)$ selects many axioms, resulting in reasonable coverage scores.
Hence, their combination results in axiom selection with sufficiently large coverage scores while maintaining a computationally feasible size.
The result is a method which solves 17.7\% more problems than the baseline, which does not employ premise selection.

%% file: results/vocab_6k/problem_statistics.tex
\begin{tabular}{llrrr}
                                             &  & Train & Validation & Test \\ \hline
Number of problems                           &  & 22179 & 2465     & 6161 \\
Average sequence length                      &  & 8.96  & 9.14     & 9.10 \\
Median sequence length                       &  & 8.00  & 8.00     & 8.00 \\
Ratio of problems containing oov tokens      &  & 0.80  & 0.85     & 0.85 \\
Ratio of problems containing only oov tokens &  & 0.05  & 0.06     & 0.07 \\ \hline
\end{tabular}

%% file: results/vocab_6k/experiment_features.tex
\begin{tabular}{llrrrrrr}
\hline
\multicolumn{2}{l}{\multirow{2}{*}{Method}} &  & \multicolumn{2}{c}{Train}                                  &  & \multicolumn{2}{c}{Validation}                             \\ \cline{4-5} \cline{7-8}
\multicolumn{2}{l}{}           &  & \multicolumn{1}{c}{Jaccard} & \multicolumn{1}{c}{Coverage} &  & \multicolumn{1}{c}{Jaccard} & \multicolumn{1}{c}{Coverage} \\ \cline{4-5} \cline{7-8} 
\multirow{3}{*}{Supervised}    & Problem    &  & \textbf{0.40}               & \textbf{0.50}                &  & \textbf{0.23}               & \textbf{0.32}                \\
                               & Premise    &  & 0.35                        & 0.46                         &  & 0.22                        & 0.31                         \\
                               & Conjecture &  & 0.20                        & 0.29                         &  & 0.14                        & 0.22                         \\ \cline{1-2} \cline{4-5} \cline{7-8} 
\multirow{3}{*}{Unsupervised}  & Problem    &  & 0.30                        & 0.40                         &  & 0.19                        & 0.27                         \\
                               & Premise    &  & 0.20                        & 0.32                         &  & 0.15                        & 0.25                         \\
                               & Conjecture &  & 0.18                        & 0.29                         &  & 0.14                        & 0.22                         \\ \hline
\end{tabular}

%% file: results/vocab_6k/experiment_order.tex
\begin{tabular}{lllllll}
\multirow{2}{*}{Axiom Order} &  & \multicolumn{2}{c}{Train}     &  & \multicolumn{2}{c}{Validation} \\ \cline{3-4} \cline{6-7} 
                             &  & Jaccard       & Coverage      &  & Jaccard        & Coverage      \\ \cline{1-1} \cline{3-4} \cline{6-7} 
Original                     &  & 0.40          & 0.50          &  & 0.23           & 0.32          \\
Length                       &  & \textbf{0.43} & \textbf{0.53} &  & \textbf{0.24}  & \textbf{0.33} \\
Frequency                    &  & 0.39          & 0.49          &  & \textbf{0.24}  & \textbf{0.33} \\
Random                       &  & 0.23          & 0.25          &  & 0.15           & 0.16          \\
Global Random                &  & 0.39          & 0.49          &  & 0.22           & 0.31          \\ \hline 
\end{tabular}

%% file: results/vocab_6k/experiment_decoding.tex
\begin{tabular}{llrrrrrrrrrrr}
                &  & \multicolumn{3}{c}{1}                                                                 & \multicolumn{1}{l}{} & \multicolumn{3}{c}{2}                                                                  & \multicolumn{1}{l}{} & \multicolumn{3}{c}{4}                                                                  \\ \cline{3-5} \cline{7-9} \cline{11-13} 
                &  & \multicolumn{1}{l}{Jaccard} & \multicolumn{1}{l}{Coverage} & \multicolumn{1}{l}{Size} & \multicolumn{1}{l}{} & \multicolumn{1}{l}{Jaccard} & \multicolumn{1}{l}{Coverage} & \multicolumn{1}{l}{Size}  & \multicolumn{1}{l}{} & \multicolumn{1}{l}{Jaccard} & \multicolumn{1}{l}{Coverage} & \multicolumn{1}{l}{Size}  \\ \hline
Greedy          &  & \textbf{0.29}               & \textbf{0.39}                & 6.71                     &                      & \textbf{0.22}               & \textbf{0.52}                & 12.82                     &                      & \textbf{0.15}               & \textbf{0.64}                & 24.25                     \\
Top-32          &  & 0.16                        & 0.30                         & 9.07                     &                      & 0.12                        & 0.43                         & 19.92                     &                      & 0.09                        & 0.58                         & 37.86                     \\
Top-64          &  & 0.16                        & 0.29                         & 9.13                     &                      & 0.11                        & 0.43                         & 20.36                     &                      & 0.08                        & 0.57                         & 39.46                     \\
Top-128         &  & 0.16                        & 0.29                         & 9.20                     &                      & 0.11                        & 0.42                         & 20.59                     &                      & 0.08                        & 0.56                         & 40.17                     \\
Top-256         &  & 0.15                        & 0.29                         & 9.21                     &                      & 0.11                        & 0.42                         & 20.64                     &                      & 0.08                        & 0.56                         & 40.40                     \\ \hline
Temperature-1.0 &  & 0.15                        & 0.28                         & 9.22                     &                      & 0.14                        & 0.38                         & 15.40                     &                      & 0.11                        & 0.50                         & 27.77                     \\
Temperature-0.9 &  & 0.17                        & 0.30                         & 8.78                     &                      & 0.16                        & 0.40                         & 14.33                     &                      & 0.12                        & 0.51                         & 25.37                     \\
Temperature-0.8 &  & 0.19                        & 0.32                         & 8.42                     &                      & 0.18                        & 0.41                         & 13.27                     &                      & 0.13                        & 0.51                         & 23.10                     \\
Temperature-0.5 &  & 0.25                        & 0.37                         & 7.49                     &                      & 0.24                        & 0.43                         & 10.60                     &                      & 0.17                        & 0.50                         & 17.15  \\
\hline
\end{tabular}